# TYPE Ia SUPERNOVA RATE IN THE GALACTIC CENTER REGION

Stéphane Schanne[1,*], Michel Cassé[1,2], Patrick Sizun[1], Bertrand Cordier[1], Jacques Paul[1]

[1] CEA-Saclay, DAPNIA/Service d'Astrophysique, F-91191 Gif sur Yvette, France
[2] Institut d'Astrophysique de Paris, 98 bis Bd Arago, F-75014, Paris, France
[*] corresponding author: schanne{at}hep.saclay.cea.fr

## ABSTRACT

According to recent analyses of the type-Ia supernova rate as a function of redshift, delayed and prompt type-Ia supernovae (SN Ia) should explode respectively in the Galactic bulge and in the nuclear bulge, a gas rich structure with ongoing star formation, located in the central region of the Milky Way. We estimate the rate of type-Ia supernovae in the Galactic bulge and nuclear bulge. We show that this rate is insufficient by an order of magnitude to explain by positron escape from type-Ia supernovae envelopes alone the large positron injection rate into the Galactic central region, as re-observed recently by the Spectrometer on INTEGRAL, which amounts to $1.25 \times 10^{43}$ e$^+$ s$^{-1}$ and would require 0.5 SN Ia explosions per century.

## 1. INTRODUCTION

The spectrometer SPI [1] on ESA's gamma-ray satellite INTEGRAL has recently presented refined measurements of the 511 keV γ-ray line emission resulting from $e^+$-$e^-$ annihilation in the Galactic center region. The 511 keV flux from this region measured by SPI to $\Phi_{511}=0.96^{+0.21}_{-0.14}\,10^{-3}$ ph/cm$^2$/s [2] is located in a narrow (~2.7 keV FWHM), non-shifted line at 511.02 ±0.09 keV. Additionally, SPI has provided constraints on the morphology of the emission region, which can be adequately described by a spherical distribution with a radial Gaussian profile and a FWHM of ~8°, while ruling out a single point source [3,4]. The extension of the annihilation region appears compatible in size and shape with the Galactic bulge. Furthermore, by measuring with SPI the orth-positronium continuum located at E<511 keV, [5] have confirmed results of earlier measurements by OSSE [6], that the dominant fraction of positrons ($f_{Ps}=0.94\pm0.06$) form a positronium intermediate state before annihilation.

From those results, the rate of $e^+$ annihilation from the region around the Galactic center, located at R=8.0±4 kpc [7], can be computed as $L_{e+}=\Phi_{511}(4\pi R^2)^{-1}(2-3 f_{Ps}/2)^{-1}$, which is the enormous amount of $1.25^{+0.35}_{-0.29}\,10^{43}$ e$^+$/s. Assuming a steady-state production-annihilation, the same amount of positrons must be injected each second into the Galactic bulge region. Consequently the question of the nature of this powerful positron source is raised.

## 2. SN Ia AS GALACTIC POSITRON SOURCE?

Among all astrophysical source candidates, up to now, supernovae of type Ia (SN Ia) have been proposed as the main source of positrons in the Galactic bulge [8]. Indeed, in the Galactic bulge, formed of very old stars (~10 Gyr), core collapse SN are not expected to take place; but those stars, if located in a binary system, could very well explode via SN Ia. Furthermore, during their explosion, SN Ia produce a copious amount of radioactive $^{56}$Ni nuclei – about 0.6 solar masses (M$_\odot$) – which decay to $^{56}$Co (after 6 days half-life), and subsequently to $^{56}$Fe (after 77 days) releasing a positron in 19% of the cases. The expanding envelope of a SN Ia is thinner than for a core-collapse SN and positrons therefore have a chance to escape from it and to be released in the interstellar medium, where they can form the observed positronium and annihilate into 511 keV γ-rays.

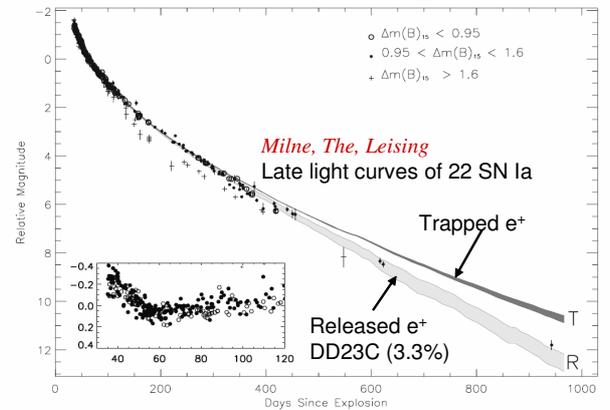

*Fig. 1. The model DD23D by Milne, The and Leising [9] of 22 SN Ia shows that a release of 3.3% of positrons is needed to describe their late light curves.*

In their model DD23C, Milne et al. [9] (Fig. 1) have predicted that 3.3% of the positrons escape a typical SN Ia envelope, reaching the total amount of $8\,10^{52}$ e$^+$ released. Firstly, in a simple calculation using the $^{56}$Ni decay law, we verify that, in order to release 3.3% of the initially produced positrons, the envelope must become transparent at latest 390 days after the explosion (in case of no mixing of the $^{56}$Ni with the ejecta), which is consistent with the timescale in Fig. 1. Since the optical light curve of a SN is powered by radioactivity (among others the produced positrons), the late light curve of typical SN Ia should have a steeper decline than by radioactive decay alone, due to the release of the positrons. To refine the model more observations of late SN Ia light curves are needed. Secondly, we deduce that, under the assumption of a steady-state production-annihilation of positrons, in order to explain the observed positron annihilation rate $L_{e+}$ by SN Ia events in the Galactic bulge alone, a mean SN Ia explosion rate of $0.50^{+0.14}_{-0.11}$ per century is required (where errors of the SN Ia model are not taken into account). But is this rate consistent with the observations?

## 3. GALACTIC BULGE SN Ia RATE ESTIMATE

Our approach to get an estimate of the SN Ia rate in the Galactic bulge is based on recent SN Ia rate measurements, obtained from studies of statistical properties of large galaxy samples, in which the galaxies are characterized by a few global parameters. If we can show that the Galactic bulge corresponds to a certain class of galaxies in this parameter space, the SN Ia rate measurement for this class of galaxies gives an estimate of the SN Ia rate in the Galactic bulge.

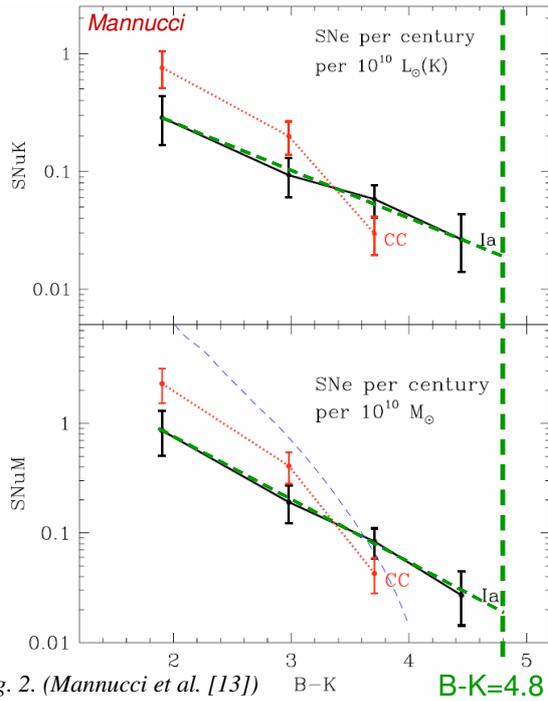

Fig. 2. (Mannucci et al. [13])

*Upper panel*: SN rate per $K$ band luminosity expressed in SNuK (number of SN per century per $10^{10}$ $L_\odot$ of $K$-band luminosity) as a function of the $B - K$ color of the parent galaxies. The thick lines are the results for type Ia (solid) and core-collapse (dotted). *Lower panel*: SN rate normalized to the stellar mass and expressed in SNuM, i.e., number of SNe per century per $10^{10}$ $M_\odot$ of stellar mass.

Our first estimate [10,11] was based on rates derived by Cappellaro et al. [12], relying on the blue integrated luminosity. In a recent paper, Mannucci et al. [13] have computed the SN Ia rate normalized to the near-infrared luminosity and to the stellar mass of the parent galaxies, using the new complete catalog of near-infrared galaxy magnitudes obtained by 2MASS. This result seems more relevant because the K-band luminosity is more directly related to the stellar population giving rise to SN Ia, namely low mass stars in binary systems. Mannucci et al. conclude that the SN Ia rate shows a sharp dependence on both the morphology and the B-K color magnitude of the parent galaxies (and therefore on the star formation activity). In particular the SN Ia rate in irregular (late type) galaxies is a factor ~20 higher than in elliptical (E/S0) galaxies. Similarly, for galaxies bluer than the color magnitude B-K=2.6 the SN Ia rate is about 30 times larger than in galaxies with B-K>4.1. Mannucci et al. show in particular (Fig 2), as a function of the B-K color of the parent galaxy, the variation of the SN Ia rate per mass unit (SNuM, expressed in numbers of SN Ia per century and per $10^{10}$ $M_\odot$ of the parent galaxy) as well as the SN Ia rate per luminosity unit (SNuK, expressed in numbers of SN Ia per century and per K-band luminosity of the parent galaxy ($L_{GAL,K}$), measured in units of $10^{10}$ K-band solar luminosity ($L_{\odot,K}$). If we determine the B-K color of the Galactic bulge, we can therefore get two estimates of the SN Ia rate in the Galactic bulge, one using the mass of the Galactic bulge, and the second using its luminosity, both of which should of course be consistent.

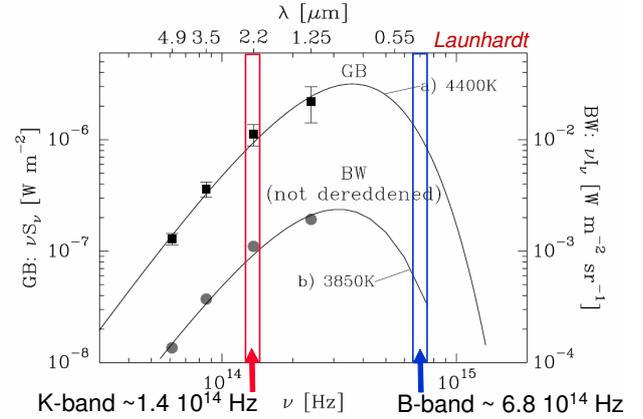

*Fig. 3. The spectral energy distribution of the Galactic bulge, from Launhardt et al. [14] is used to derive the B-K color magnitude of the Galactic bulge.*

Launhardt et al. [14] present a measurement of the spectral energy distribution of the Galactic bulge (see Fig. 3), which is best fit by an effective black-body spectrum, whose temperature is $T_{GB}$~4400 K. The B-band spectral window is defined around the frequency $\nu$=6.8 $10^{14}$ Hz and the K-band around $\nu$=1.4 $10^{14}$ Hz [15]. From [14] the integrated flux density $F_\nu$ of the Galactic bulge in the B-band is measured to be $\nu F_\nu(B)$=3.6 $10^{-7}$ W/m$^2$, and in the K-band $\nu F_\nu(K)$=9.2 $10^{-7}$ W/m$^2$. Therefore we can compute the difference of the magnitudes in the B- and K-band of the Galactic bulge, using the definition of the magnitude $m_v=2.5 \log_{10}(F_v(0)/F_v(m_v))$ as a function of the absolute flux density $F_v(m_v)$; for which the values at magnitude m=0 are $\nu F_v(0)=2.9 \ 10^{-8}$ W/m$^2$ in the B-band, and $\nu F_v(0)=9.2 \ 10^{-10}$ W/m$^2$ in the K-band [15]. As a conclusion we compute a B-K color (difference $m_B-m_K$) for the Galactic bulge of B-K=+4.8. It is interesting to notice that the Galactic bulge appears to be similar in B-K color to the sample of elliptical galaxies (E/S0) for which [14] derive a color B-K>4.1. Additionally taking into account the morphology of the Galactic bulge, and the fact that it is composed of very old stars (~10 Gyr), we would like to emphasize that the Galactic bulge can be viewed in our context as an elliptical galaxy embedded in the spiral Milky Way.

From the analysis of Mannucci et al. [13], we obtain with B-K=+4.8 for the Galactic bulge an estimate of the SN Ia rate per unit mass of 0.02±0.01 SNuM and a SN Ia rate per unit luminosity of 0.02±0.01 SNuK (dashed lines on Fig. 2; we notice again, that those numbers agree with the rate estimates of 0.044±0.015 SNuM and 0.035±0.013 SNuK obtained by [13] for their sample of elliptical galaxies).

There are several estimations of the stellar mass content of the Galactic bulge in the literature. A photometric determination of this mass has been given by [16] to be 1.3±0.5 $10^{10}$ M$_\odot$, from the Galactic bulge dynamics [17] and [18] obtain a mass of ~$10^{10}$ M$_\odot$, and a model of the old population stars in the outer Galactic bulge by [19] uses a mass of 2.03±0.26 $10^{10}$ M$_\odot$. If we assume therefore a Galactic bulge mass of 1.5±0.5 $10^{10}$ solar masses, we conclude that the SN Ia rate in the Galactic bulge is 0.03±0.02 per century.

Concerning the measurements of the luminosity of the Galactic bulge, [14] have estimated a value $L_{GB}$=1.0±0.3 $10^{10}$ L$_\odot$, while [16] determined from COBE measurements a value of 0.53±0.16 $10^{10}$ L$_\odot$ and older measurements by [20] present values of 2 $10^{10}$ L$_\odot$. Those values can be converted into units of K-band luminosity, using the translation factor $L_{GB,K}/L_{\odot,K}$=1.815 $L_{GB}/L_\odot$, which we computed using the ratio $L_{\odot,K}/L_\odot$ between the integral of the surface brightness in the K-band and the total emittance of a black body at the temperature of the Sun ($T_\odot$=5780 K), as well as the same ratio $L_{GB,K}/L_{GB}$ for the black body representing the Galactic bulge ($T_{GB}$=4400 K). For the Galactic bulge luminosity determined by [14], we conclude therefore that the SN Ia rate in the Galactic bulge is 0.036±0.021 per century, well in agreement with the value computed using the mass of the Galactic bulge as an input.

Our estimate of the rate of SN Ia explosions in the Galactic bulge of 0.03±0.02 per century, is well in agreement with the value of 0.05 predicted by Matteucci et al. [21] (Fig 6 in their paper) based on their model of the chemical evolution of the Galaxy.

This SN Ia rate is however incompatible with the hypothesis that SN Ia be the dominant Galactic positron injectors. But there could be an additional SN Ia component in the Galactic central regions, coming from explosions in the Nuclear Bulge. Are they sufficient to fill the gap?

## 4. NUCLEAR BULGE SN Ia RATE ESTIMATE

In a recent paper, Mannucci et al. [22] have discovered two populations of SN Ia which differ in the delay between the formation of the progenitor and the SN explosion, and suggest that two modes of SN Ia production are at work in the universe, a prompt and a delayed one. The delayed mode is the "standard" scenario, associated with a single degenerate system (White Dwarf) accreting matter from a companion star until the explosion is triggered after a delay of several Gyr, whereas the prompt mode may be associated with the merging of two White Dwarfs (double degenerate system), triggering the explosion after only ~$10^8$ yr. While the delayed mode certainly applies to the old (passive) stellar population of the Galactic bulge, where star formation is over since about 10 Gyr, the prompt mode may be at work in the nuclear bulge of the Galaxy, and we investigate in the following the SN Ia rate associated to this region.

The nuclear bulge (NB) of the Galaxy appears as a massive disk-like complex of stars and molecular clouds of radius ~250 pc and height ~50 pc, located in the Galactic center region, and which is distinct from the Galactic bulge, as described in Launhard et al [14]. The NB is a very active region in the Galaxy with ongoing star formation. Its total stellar mass and luminosity are 0.14±0.06 $10^{10}$ M$_\odot$ and 0.25±0.1 $10^{10}$ L$_\odot$ respectively, most of the luminosity being due to young massive main-sequence stars.

Moreover, the specific star formation rate of the nuclear bulge is comparable to the one of the whole Milky Way (for which stars are formed at a rate of approximately 0.04 new M$_\odot$ of stars per M$_\odot$ and per Gyr). Therefore in a first approach, we can consider the nuclear bulge as a spiral galaxy and estimate its SN Ia using the SN Ia rate per unit mass for spirals of 0.17±0.07 SNuM form [13]. In this way, we would get a SN Ia rate in the nuclear bulge of 0.024±0.014 per century. However, the initial mass function in the NB follows the relation $\Psi(M) \propto M^{-1.7 \text{ to } -1.9}$ [14], at variance with the standard Salpeter one, i.e. $\Psi(M) \propto M^{-2.3}$. Therefore in the nuclear bulge the number of White Dwarfs (WD) formed is enhanced, increasing the number of expected SN Ia by the same factor. Since WD form among 3 to 8 M$_\odot$ stars out of all stars formed, the number of WD per M$_\odot$ is:

$$N_{WD} = \int_3^8 \Psi(M)dM \bigg/ \int_{0.1}^{120} M\Psi(M)dM$$

In the Milky Way, this evaluates to $N_{WD}$=0.021 $M_\odot^{-1}$, while in the nuclear bulge we obtain $N_{WD}$=0.029 $M_\odot^{-1}$. Taking this enhancement into account, we estimate the rate of SN Ia explosions in the Galactic nuclear bulge to be 0.03±0.02 per century.

## 5. CROSSCHECK OF SN Ia RATES

In this section we would like to present some alternative evaluations and crosscheck them with the previously determined SN Ia rates. Indeed, there is an other way to obtain the SN Ia rate in the nuclear bulge. By comparing the type Ia SN rate (SNR) and the star formation rate (SFR) at various redshifts, and assuming that it applies locally, namely in the nuclear bulge, one can deduce the SNR from the SFR there. The SFR can be readily estimated dividing the total mass in stars in the system (1.4 $10^9$ $M_\odot$ [14]) by its age (10 Gyr), assuming a more or less constant rate of star formation (Figer et al. [23]). A short delay time between star formation and SN Ia explosion implies that the cosmic SNR follows closely the SFR. The conversion factor between SNR and SFR can be directly determined from observations. Barris and Tonry [24] obtain 1.7 $10^{-3}$ SN Ia per $M_\odot$. Thus a SN Ia rate of 0.024 per century is deduced in the nuclear bulge, which compares favorably with our previous estimate. A more extended discussion will be given in a forthcoming paper, analyzing the validity of the assumptions and the uncertainty on the SFR history due to debated corrections for dust extinction [25].

The two, prompt and delayed, SN Ia modes presented by Mannucci et al. [22] have recently been modeled by Scannapieco et al [26], giving an estimate of the rate of SN Ia as a function of the parent galaxy's mass $M$ and current star formation rate $SFR$, with parameters $A$=0.044±0.015 and $B$=2.6±1.1:

$$\frac{R(SNIa)}{100\text{yr}} = A \frac{M}{10^{10} M_\odot} + B \frac{SFR}{10^{10} M_\odot \text{ Gyr}^{-1}}$$

By applying this model to the Galactic bulge and the nuclear bulge respectively, and using (expressed in the units given in this equation) $M_{GB}$=1.5±0.5 and $SFR_{GB}$=0, and $M_{NB}$=0.14±0.06 and $SFR_{NB}$= $M_{NB}$/(10 Gyr), we obtain $R_{GB}$=0.06±0.03 and $R_{NB}$=0.04±0.03 type Ia supernovae per century, which are again compatible with our previous estimations.

## 6. CONCLUSIONS

Based on the previously exposed arguments, we have obtained an estimate of the rate of SN Ia explosions (among the old stellar population) in the Galactic bulge of 0.03±0.02 per century, and (among the young stellar population) in the Galactic nuclear bulge of equally 0.03±0.02 per century. The summed SN Ia rate of both contributions is however incompatible by a factor of about 10 with the rate of $0.50^{+0.14}_{-0.11}$ SN Ia per century required, under the hypothesis that SN Ia were the main injectors of positrons in the Galactic bulge. Even within several error bars (which must be understood here as plausible parameter ranges) we can not reconcile both numbers, and therefore we conclude that only a small fraction of SN Ia can contribute to the source of positrons in the Galactic bulge.

Since we have not displayed any errors in the SN Ia model, one could of course object that SN Ia could release up to 30% (instead of ~3%) of the positrons produced in the explosion, in order to gain the missing factor of 10. This huge escape fraction could either occur in models where the radioactive isotopes are well mixed with the ejecta in the outer layer of the envelope, resulting in an early escape of the positrons (at latest 143 instead of 390 days after the explosion). In this case, at the epoch of transparency, a significant amount of energy would be lost by the envelope, taken away by the escaping positrons which are produced in the radioactive $^{56}$Co decay with a spectrum peaking around 640 keV (the 847 keV photons produced in the same decay would escape even earlier). This energy loss would result in a less efficient heating of the ejecta, and therefore a steeper decline in the SN Ia light curve related to the moment of transparency to positrons would be expected, which seems not to be observed, see e.g. [27] or [28] who report observation of the light curve of SN 1996x for about 500 days; Note that the late light curve of SN2000cx is consistent with no positron escape at all, but this SN Ia is rather peculiar spectroscopically [29], i.e. superficially, which does not prove, however, that it is abnormal in its deeper structure. Of course the escape fraction of the positrons from the SN Ia envelope remains to be confirmed by further observations of SN Ia light curves at late time and more sophisticated (2D) modeling of SN Ia explosions (see e.g. Röpke, Hillebrandt et al 2005 [30]).

This result on the SN Ia deficiency to inject sufficient positrons into the Galactic bulge opens the door to other explanations, which are numerous. While classical core collapse SN have a too thick envelope to release positrons from $^{56}$Co, asymmetric explosions like hypernovae [10,11] could produce copious amounts of $e^+$. However in the steady-state hypothesis the hypernova rate in the Galactic center region remains probably too low, even if a recent starburst in the region of the Galactic nucleus could alleviate this problem. The contribution form other radioactive isotopes (like $^{26}$Al from massive stars, $^{44}$Ti from supernovae, and $^{22}$Na from novae) is also too to low to close the gap. Other astrophysical source candidates are compact objects – among which low-mass X-ray binaries [31] which are more numerous in the Galactic central regions but for which the positron production is unknown – and microquasar jets which are however inactive during long periods of time. The positrons could also be the remnant of past gamma-ray burst(s) in the Galaxy [32].

In the absence of a major astrophysical candidate source for the Galactic bulge positrons, other processes have been proposed, among which the annihilation into $e^+$–$e^-$ pairs of a new elementary particle, candidate for the observed dark matter in the universe, which could have escaped detection at past accelerator experiments because of its low mass [33,34,35]

Further long exposure observations by SPI aboard INTEGRAL will help, by a more precise mapping of the positron annihilation region (especially the determination of the 511 keV Galactic disk component) and flux determination, to get more stringent constraints on the nature of the Galactic positron source.

## 7. ACKNOWLEDGMENTS

The authors would like to thank their colleagues of SPI, the Spectrometer aboard the European Space Agency's gamma-ray space telescope INTEGRAL, for stimulating discussions.